\documentclass{PoS}
\usepackage{amsmath}
\usepackage{amssymb}
\usepackage{graphicx}
\usepackage{bbold}
\newcommand{\hc}{\hbox {h.c.}}
\renewcommand{\Re}{\mbox{Re\thinspace}}
\renewcommand{\Im}{\mbox{Im\thinspace}}

\title{CP violation in extended Higgs sectors}

\ShortTitle{CP violation}

\author{O. M. Ogreid\\
        Western Norway University of Applied Sciences, \\
Postboks 7030, N-5020 Bergen, Norway, \\
        E-mail: \email{omo@hvl.no}}

\author{\speaker{P. Osland} 
\\
Department of Physics and Technology, University of Bergen, \\
Postboks 7803, N-5020  Bergen, Norway,\\
        E-mail: \email{Per.Osland@uib.no}}

\author{M. N. Rebelo\\
        Centro de F\'isica Te\'orica de Part\'iculas -- CFTP and Dept de F\' \i sica\\
Instituto Superior T\'ecnico -- IST, Universidade de Lisboa, Av. Rovisco Pais, \\
P-1049-001 Lisboa, Portugal\\
        E-mail: \email{rebelo@tecnico.ulisboa.pt}}

\abstract{We discuss how one can identify CP violation (and conservation) in
multi-Higgs-doublet potentials. After a brief review of CP violation
in the 2HDM, we refer to the fact that for NHDM with $N \geq 3$ the well
known methods useful in the case  $N = 2$ have not been generalized
in order to provide a set of well defined necessary and sufficient
conditions for CP conservation. We then present a simple method,
proposed by the authors, to be used in such cases.
Two non-trivial examples based on an $S_3$-symmetric three-doublet model
are analyzed by means of this new method.}

\FullConference{Corfu Summer Institute 2017 'School and Workshops on Elementary Particle Physics and Gravity'\\
		2-28 September 2017\\
		Corfu, Greece}

\begin{document}

\section{Introduction}

Multi-Higgs-doublet models provide a rich framework for Beyond-Standard-Model physics. Models with two doublets (2HDM) have long received a lot of attention \cite{Gunion:1989we,Branco:2011iw}.
Models with three Higgs doublets are also well motivated and for a recent review of such models see, for instance, Ref.~\cite{Ivanov:2017dad}. 
As the number of doublets increases, so does the number of free parameters.
Symmetries play an important r\^ ole in reducing this number, thus adding predictive power to the models.

The most general NHDM potential is given by
\begin{equation} \label{Eq:pot3hdm}
V=Y_{ab}\Phi_{a}^\dagger\Phi_b
+\frac{1}{2}Z_{abcd}(\Phi_{a}^\dagger\Phi_b)(\Phi_{c}^\dagger\Phi_d)
\end{equation}
where the $\Phi$ are SU(2) doublets and
$a,b,c,d$ run over the values 1 to $N$, with repeated indices to be summed over.
For $N=3$ the potential has 3 diagonal bilinear terms (real): $Y_{11}$, $Y_{22}$, $Y_{33}$
and 3 off-diagonal ones (complex) satisfying $Y_{ba}^\ast=Y_{ab}$.
Furthermore, there are many $Z_{abcd}$, some real, some complex, satisfying:
\begin{equation}
Z_{badc}^\ast=Z_{abcd}, \quad Z_{cdab}=Z_{abcd}.
\end{equation}
All counted, there are 54 parameters, but they are not all independent, since
we may rotate:
\begin{equation}
\begin{pmatrix}
\Phi_1^\prime \\
\Phi_2^\prime \\
\Phi_3^\prime
\end{pmatrix}
=U
\begin{pmatrix}
\Phi_1 \\
\Phi_2 \\
\Phi_3
\end{pmatrix}
\end{equation}
with $U$ an arbitrary unitary matrix. This way, one may diagonalize the bilinear part (removing 6 parameters) and also remove 2 relative phases between the $\Phi$. The remaining number of linearly independent parameters is thus 46 \cite{Olaussen:2010aq}. For comparison, the number of independent parameters in the 2HDM is 11.
This illustrates the fact that the number of free parameters in multi-Higgs-doublet 
models grows fast with the number of doublets, thus leading to a very rich structure. Without further
constraints these models may lead, for instance, to potentially dangerous flavour changing neutral currents in the quark sector, and 
may also lead to other new phenomena already ruled out by experiment.

We will discuss methods for identifying CP violation or conservation, in $N\geq3$ models, 
and will illustrate their power by applying them to an $S_3$-symmetric, ten-parameter $N=3$ potential.
These methods can equally well be applied to the case of $N=2$. However, as will be explained in section
3,  this case is simple and has already been well studied. \\

\section{Identifying CP violation}

CP violation in gauge theories requires the introduction of a scalar
sector. Pure gauge theories including fermions cannot violate CP \cite{Grimus:1995zi}.

If all coefficients in the potential and all vacuum expectation values (vevs) are real, 
then CP is  conserved. However, the converse does not always hold. For instance, it 
is possible to have a complex potential that does not violate CP explicitly, and it 
is also possible to have explicit CP conservation with a real potential together with 
non-trivial complex vevs that do not violate CP spontaneously.

The most general CP transformation for multi-Higgs models is given by 
\cite{Branco:1983tn, Branco:1999fs}:
\begin{equation}
\Phi_i \stackrel{\text{CP}}{\longrightarrow} W_{ij} \Phi^\ast_j
\label{abc}
\end{equation}
for $W$ a unitary, arbitrary transformation. Eq.~(\ref{abc}) is a combination of the CP
transformation of each single Higgs doublet with a Higgs basis transformation.
This, or in alternative the fact that physics does not change with a change 
of Higgs basis, has been  exploited in the identification of CP-odd 
re-parametrization invariants sensitive to explicit CP violation 
\cite{Branco:2005em,Gunion:2005ja}. For the 2HDM these conditions are equivalent to conditions 
written in terms of the charged-Higgs mass and its quartic coupling \cite{Grzadkowski:2016szj}.
If all coefficients of the potential are real, then there is no explicit CP violation. 
Assuming that the potential is real, there is still the possibility of 
having spontaneous CP violation through phases appearing in the vevs.  CP violation can 
only be considered to be spontaneous if there is explicit CP conservation. 

In the case of the 2HDM a full set of invariant conditions sensitive to 
spontaneous CP violation involving the potential and the vevs has been derived
\cite{Lavoura:1994fv,Botella:1994cs}.  For the 2HDM, Higgs-basis-invariant
conditions can be expressed in terms of masses and couplings 
\cite{Lavoura:1994fv,Botella:1994cs,Grzadkowski:2014ada,Grzadkowski:2015zma}. 
The technique to generate such invariants can be applied in theories with more 
than two Higgs doublets. However, a full set of necessary and sufficient conditions 
for CP conservation in the cases of $N=3$ or higher
has not yet been identified. There are examples in the literature for special cases 
with particular symmetries \cite{Varzielas:2016zjc,deMedeirosVarzielas:2017ote}.
Higgs basis invariants have also been applied to determine tree level Higgs couplings and masses
without putting the emphasis on CP violation \cite{Davidson:2005cw}.
 
An alternative method to determine whether or not there is spontaneous CP violation in 
an NHDM was provided in Ref.~\cite{Branco:1983tn}  where it was shown
that if there is a symmetry $U$ of the Lagrangian, acting on the Higgs doublets,
${\cal L} (U  \Phi ) = {\cal L} (\Phi) $, 
under which the  Higgs vevs satisfy the relation: 
\begin{equation} \label{Eq:U-spont}
 U_{ij} \langle 0| \Phi_j |0\rangle^\ast = \langle 0| \Phi_i |0\rangle
\end{equation}
then the vacuum is invariant and there is no spontaneous CP violation. This is
a very powerful relation but in some cases finding the matrix $U$ that satisfies 
Eq.~(\ref{Eq:U-spont}) may not be straightforward.

For more complicated cases a simple procedure to determine whether or not there
is CP violation has been proposed in \cite{Ogreid:2017alh} and
consists of starting by making a transformation to a Higgs
basis where only one of the Higgs doublets acquires a vev different from zero,
chosen to be real, and all other doublets have zero vevs  \cite{Donoghue:1978cj,Georgi:1978ri}. 
The next step consists in using the freedom to rephase the 
doublets with zero vev or, if necessary, even to perform a $U(N-1)$ unitary 
rotation of these fields in order to make all the coefficients of the potential real.
This method is sensitive to both explicit and spontaneous CP violation. 

In section 4 we will show how this method can be used in the case of an $S_3$-symmetric, 
ten-parameter $N=3$ potential, where the determination of the matrix $U$ of 
Eq.~(\ref{Eq:U-spont}) is not straightforward.

\section{Review of CP violation in the 2HDM}

The most general 2HDM contains three neutral physical scalars, but these need not be eigenstates of CP. 
This can be illustrated by going to the Higgs basis \cite{Donoghue:1978cj,Georgi:1978ri} where 
the two Higgs doublets are parametrized as
\begin{equation}
\Phi_1=\left(
\begin{array}{c}G^+\\ (v+\eta_1+iG_0)/\sqrt{2}
\end{array}\right), \quad
\Phi_2=\left(
\begin{array}{c}H^+\\ (\eta_2+i\chi_2)/\sqrt{2}
\end{array}\right),
\label{vevs}
\end{equation}
and the potential takes the form
\begin{align}
\label{Eq:pot}
V(\Phi_1,\Phi_2) &= -\frac12\left\{m_{11}^2\Phi_1^\dagger\Phi_1
+ m_{22}^2\Phi_2^\dagger\Phi_2 + \left[m_{12}^2 \Phi_1^\dagger \Phi_2
+ \hc\right]\right\} \nonumber \\
& + \frac{\lambda_1}{2}(\Phi_1^\dagger\Phi_1)^2
+ \frac{\lambda_2}{2}(\Phi_2^\dagger\Phi_2)^2
+ \lambda_3(\Phi_1^\dagger\Phi_1)(\Phi_2^\dagger\Phi_2) 
+ \lambda_4(\Phi_1^\dagger\Phi_2)(\Phi_2^\dagger\Phi_1)\nonumber \\
&+ \frac12\left[\lambda_5(\Phi_1^\dagger\Phi_2)^2 + \hc\right]
+\left\{\left[\lambda_6(\Phi_1^\dagger\Phi_1)+\lambda_7
(\Phi_2^\dagger\Phi_2)\right](\Phi_1^\dagger\Phi_2)
+{\rm \hc}\right\}.
\end{align}
The parameters $m_{12}^2$ together with $\lambda_5$, $\lambda_6$ and $\lambda_7$ can be complex.
The minimization conditions impose $m_{12}^2 = v^2 \lambda_6$.
The mass-squared matrix will depend on the parameters $m_{22}^2$, $\lambda_1$, $\lambda_3$, $\lambda_4$, $\lambda_5$ and $\lambda_6$ :
\begin{equation}
{\cal M}^2=
\begin{pmatrix}
	\lambda_1 v^2 & \Re\lambda_6 v^2 & -\Im\lambda_6 v^2 \\
	\Re\lambda_6 v^2 & \frac{1}{2}\left(-m_{22}^2+(\lambda_3+\lambda_4+\Re\lambda_5)v^2\right) & -\frac{1}{2}\Im\lambda_5v^2 \\
	-\Im\lambda_6 v^2 & -\frac{1}{2}\Im\lambda_5v^2 & \frac{1}{2}\left(-m_{22}^2+(\lambda_3+\lambda_4-\Re\lambda_5)v^2\right)
\end{pmatrix}.
\end{equation}
Notice that the mass matrix does not include $\lambda_7$. If $\lambda_5$ and  $\lambda_6$ can 
be made simultaneously real by a redefinition of $\Phi_2$ there is no mixing among CP-even and 
CP-odd fields. This is the case,
for instance, if $\lambda_6=0$ or $\lambda_5=0$. In particular, for $\lambda_6=0$ 
we can make $\lambda_5$ real by rephasing $\Phi_2$ and the mass matrix becomes automatically diagonal.
However, in order to conclude that CP is conserved one must check whether or not $\lambda_7$ can also 
be made real with the same rephasing of $\Phi_2$ that makes $\lambda_5$ and  $\lambda_6$ real, 
otherwise there will be CP violation in the trilinear and quartic couplings.

These conditions will look different in a general (non-Higgs) basis, but the different 
possibilities of having CP conservation or violation can be sorted out by exploring the 
basis-transformation invariants mentioned above, see  \cite{Lavoura:1994fv,Botella:1994cs} 
and \cite{Branco:2005em,Gunion:2005ja}.

A different approach is to ask whether a basis exists in which the potential and the vevs 
are simultaneously real \cite{Gunion:2002zf}. When applied in the Higgs basis 
\cite{Ogreid:2017alh} this constitutes a powerful
test for the study of  multi-Higgs-doublet models as illustrated in what follows.

\section{The use of the Higgs basis to test for CP Conservation}

\subsection{The 2HDM with real coefficients}
Without loss of generality, the 2HDM potential can be written as:\footnote{Here, we follow the notation of ref.~\cite{Lee:1973iz}, the $\lambda_1$ and $\lambda_2$ should not be confused with those of Eq.~(\ref{Eq:pot}).}
\begin{align}
V(\phi)
&=-\lambda_1\phi_1^\dagger\phi_1-\lambda_2\phi_2^\dagger\phi_2 \nonumber \\
&+A(\phi_1^\dagger\phi_1)^2+B(\phi_2^\dagger\phi_2)^2+C(\phi_1^\dagger\phi_1)(\phi_2^\dagger\phi_2) 
+\bar C (\phi_1^\dagger\phi_2)(\phi_2^\dagger\phi_1) 
\nonumber \\
&+\frac{1}{2}[
(\phi_1^\dagger\phi_2)(D\phi_1^\dagger\phi_2+E\phi_1^\dagger\phi_1+F\phi_2^\dagger\phi_2) +\text{h.c.}],
\label{asd}
\end{align}
with the bilinear part already diagonal, making use of the freedom in the choice of
Higgs basis as described in the Introduction. In general, the coefficients $D$, $E$ and $F$
can all be complex, but at least one of then can be made real by simply rephasing 
one of the doublets. In this way one ends up with the eleven independent real 
parameters mentioned before. There is explicit CP violation if it is not possible 
to make all three coefficients real at the same time. 

It is well known that, even with real coefficients, the 2HDM can violate CP spontaneously  
\cite{Lee:1973iz}, since, in this case the potential allows for non-real vevs: 
$(\rho_1e^{i \theta}, \rho_2)$.  
In the absence of an additional symmetry of the potencial and for a nontrivial phase, there 
is no way of verifying  the condition given by  Eq.~(\ref{Eq:U-spont}). 
If a $Z_2$ symmetry is imposed on the 2HDM Lagrangian there is neither explicit nor spontaneous 
CP violation. The possibility of spontaneous CP violation in two-doublet models with a 
softly broken discrete symmetry was pointed out in \cite{Branco:1985aq}.

Starting with the potential of the 2HDM in the notation of Eq.~(\ref{asd}) and with real coefficients 
there is still the possibility of having spontaneous CP violation \cite{Lee:1973iz} since, as mentioned
above, there is a region of parameters where the vevs are of the form $(\rho_1e^{i \theta}, \rho_2)$, 
with $\theta$ non trivial and both $\rho_1$ and $\rho_2$ different from zero. 

The Higgs basis is reached via the transformation
\begin{equation}
\left( \begin{array}{c} 
\phi_1^\prime \\
\phi_2^\prime \\
\end{array}  \right) = \frac{1}{v} \left( \begin{array}{cc}
1 & 0  \\
0 & e^{i \chi} \\
\end{array} \right) 
\left( \begin{array}{cc} 
\rho_1 & \rho_2 \\
- \rho_2 & \rho_1 \\ 
\end{array} \right) \left( \begin{array}{cc} 
e^{- i \theta} & 0  \\
0 & 1 \\
\end{array} \right) 
 \left( \begin{array}{c}
\phi_1 \\
\phi_2 \\
\end{array}  \right)
\label{basis}
\end{equation}
with (normalization) $v^2 = \rho_1^2+\rho_2^2$. This basis is defined as the basis where the
vevs are real and are of the form $(v, 0)$. In this basis CP violation manifests itself by the
impossibility of making the coefficients of the potential real by rephasing $\Phi_2$.

This transformation generates off-diagonal terms $\phi_1^\dagger\phi_2$ and $\phi_2^\dagger\phi_1$ with complex coefficients given by:
\begin{eqnarray}	
\frac{(\lambda_1-\lambda_2)\rho_1\rho_2\, e^{\pm i\chi}}{v^2}.
\end{eqnarray}
The coefficients of these bilinear terms are only real if $\sin \chi = 0$ or $\lambda_1 = \lambda_2$.
In each case, requiring the quartic part of the potential to be also real imposes new constraints on the 
parameters of the potential. Therefore, confirming in this way, that in general CP is not conserved even if we 
start with a real 2HDM potential.

\subsection{Two non-trivial 3HDM cases}

In order to illustrate the difficulty that may arise in finding a matrix $U$ that satisfies 
Eq.~(\ref{Eq:U-spont}) in a multi-doublet theory, consider the following $S_3$-symmetric 
3HDM potential:
\begin{equation}
 V= V_2+V_4
 \end{equation}
with \cite{Derman:1978rx,Das:2014fea}
\begin{subequations} \label{Eq:V-S_3}
\begin{align}
V_2&=\mu_0^2 h_S^\dagger h_S +\mu_1^2(h_1^\dagger h_1 + h_2^\dagger h_2), \\
V_4&=
\lambda_1(h_1^\dagger h_1 + h_2^\dagger h_2)^2 
+\lambda_2(h_1^\dagger h_2 - h_2^\dagger h_1)^2
+\lambda_3[(h_1^\dagger h_1 - h_2^\dagger h_2)^2+(h_1^\dagger h_2 + h_2^\dagger h_1)^2]
\nonumber \\
&+ \lambda_4[(h_S^\dagger h_1)(h_1^\dagger h_2+h_2^\dagger h_1)
+(h_S^\dagger h_2)(h_1^\dagger h_1-h_2^\dagger h_2)+\hc] 
+\lambda_5(h_S^\dagger h_S)(h_1^\dagger h_1 + h_2^\dagger h_2) \nonumber \\
&+\lambda_6[(h_S^\dagger h_1)(h_1^\dagger h_S)+(h_S^\dagger h_2)(h_2^\dagger h_S)] 
+\lambda_7[(h_S^\dagger h_1)(h_S^\dagger h_1) + (h_S^\dagger h_2)(h_S^\dagger h_2) +\hc]
\nonumber \\
&+\lambda_8(h_S^\dagger h_S)^2,
\end{align}
\end{subequations}
where all parameters are real.
Here, $h_1$ and $h_2$ are members of an $S_3$ doublet, whereas $h_S$ is a singlet. These
fields are related to the fields in the defining representation of $S_3$, $\phi_i, i =1, 2, 3$, by:
\begin{equation} \label{Eq:doublet}
\left(
\begin{array}{c}h_1\\ h_2
\end{array}
\right)
=\left(
\begin{array}{c}\frac{1}{\sqrt2}(\phi_1-\phi_2)\\ \frac{1}{\sqrt6}(\phi_1+\phi_2-2\phi_3)
\end{array}
\right), \quad
h_S=\frac{1}{\sqrt3}(\phi_1+\phi_2+\phi_3).
\end{equation}
The $SU(2)$ structure is decomposed as
\begin{equation}
h_k=\begin{pmatrix} h_k^+ \\ (w_k+\eta_k+i\chi_k)/\sqrt2
\end{pmatrix}, \quad k=1,2,S.
\end{equation}
In addition to 11 distinct real vacua,
this potential possesses 14 distinct complex vacua \cite{Emmanuel-Costa:2016vej}, where at least one of the parameters
\begin{equation}
(w_1, w_2, w_S)
\end{equation}
is complex.
We shall discuss two particular cases, where some vacuum expectation values are complex, and yet, CP is conserved.
These will in the following be denoted C-III-c and C-IV-e \cite{Emmanuel-Costa:2016vej}, where the ``III'' and ``IV'' refer to the number of consistency constraints that are required for the particular vacuum.
\subsubsection{The vacuum C-III-c}
This vacuum is characterized by
\begin{equation}
(w_1, w_2, w_S)=(\hat w_1e^{i\sigma_1}, \hat w_2e^{i\sigma_2}, 0),
\end{equation}
where $\hat w_1$ and $\hat w_2$ are real, and the three constraints are
\begin{gather}
\mu_1^2=-(\lambda_1+\lambda_3)(\hat{w}_1^2+\hat{w}_2^2),\\
\lambda_2+\lambda_3=0, \quad \lambda_4=0.
\end{gather}

It is not apparent, at first sight, that this vacuum conserves CP,
mainly due to the fact that  $\hat w_1$ and $\hat w_2$ are different in general.
It is possible to show that CP is conserved \cite{Ogreid:2017alh}, by first making a transformation to the 
Higgs basis \cite{Donoghue:1978cj,Georgi:1978ri}:
\begin{equation}
\left( \begin{array}{c}
h_1^\prime \\
h_2^\prime \\
h_S^\prime \\
\end{array}  \right) = \frac{1}{v} \left( \begin{array}{ccc} 
\hat {w_1}  & \hat {w_2} & 0 \\
 \hat {w_2} & - \hat {w_1}& 0 \\ 
 0 & 0 & v \\
\end{array} \right) \left( \begin{array}{ccc} 
e^{-i\sigma_1} & 0 & 0 \\
0 & e^{-i\sigma_2} & 0 \\
0 & 0 & 1 
\end{array} \right) 
 \left( \begin{array}{c}
h_1 \\
h_2 \\
h_S \\
\end{array}  \right)
\label{basis2}
\end{equation}
with the normalization given by
$v^2= (\hat {w_1}^2 + \hat {w_1}^2)$.
The coefficients of the potential remain real, so we see explicitly that CP is conserved.

Finding a transformation $U$ satisfying Eq.~(\ref{Eq:U-spont}) is now possible by
exploiting the fact that $\hat w_S=0$. We construct a transformation of the following form
\begin{equation}
U = 
e^{i(\delta_1+\delta_2)}
\begin{pmatrix}
\cos\theta &\sin\theta & 0\\
-\sin\theta & \cos\theta & 0 \\
0 & 0 & 1
\end{pmatrix}
\begin{pmatrix}
0 & 1 & 0 \\
1 & 0 & 0 \\
0 & 0 & 1
\end{pmatrix}
\begin{pmatrix}
\cos\theta &-\sin\theta & 0 \\
\sin\theta & \cos\theta & 0 \\
0 & 0 & 1
\end{pmatrix}
\label{u2}
\end{equation}
and choose $\theta$ such that the vevs become \cite{Ogreid:2017alh}
\begin{equation}
(ae^{i\delta_1},ae^{i\delta_2},0),
\end{equation}
i.e., the non-zero vevs have the same modulus. This step is based on the fact
that for $\lambda_4=0$ the potential acquires an $SO(2)$ symmetry between 
$h_1$ and $h_2$.
Now, an overall rephasing of the fields, which also leaves the potential invariant, given by
$e^{-i(\delta_1+\delta_2)/2}$ 
leads to vevs of the form
\begin{equation}
(ae^{i\delta},ae^{-i\delta},0).
\end{equation}
This overall phase rotation is not felt by the vev of $h_S$ because it is zero.
Now, due to $\lambda_4=0$, the potential is also symmetric under
\begin{equation}
h_1 \leftrightarrow h_2,
\end{equation}
and one can see that CP is conserved because Eq.~(\ref{Eq:U-spont}) can be
verified by means of the matrix $U$ written above, which encodes the three steps
just described. Each one of these steps is based on a symmetry property of the potential.

\subsubsection{The vacuum C-IV-e}
This case differs significantly from the previous one, since
$\hat w_S \neq 0$. However, $\hat w_1$ and $\hat w_2$ are now related.
In fact, the vacuum is given by
\begin{equation} \label{Eq:C-IV-e-vacuum}
\left(\sqrt{-\frac{\sin 2\sigma_2}{\sin 2\sigma_1}}\hat w_2e^{i\sigma_1},
\hat w_2e^{i\sigma_2},\hat w_S\right),
\end{equation}
subject to the four constraints
\begin{gather}
\mu _0^2=\frac{ \sin ^2\left(2 \left(\sigma _1-\sigma _2\right)\right)}{ \sin ^2\left(2 \sigma _1\right)}\left(\lambda _2+\lambda _3\right) \frac{\hat{w}_2^4}{\hat{w}_S^2}
 -\frac{1}{2}\left(1-\frac{\sin 2 \sigma _2}{ \sin 2 \sigma _1}\right) \left(\lambda _5+\lambda _6\right) \hat{w}_2^2-\lambda _8 \hat{w}_S^2, \\
\mu _1^2= - \left(1-\frac{\sin 2 \sigma _2}{ \sin 2 \sigma _1}\right) \left(\lambda _1-\lambda _2\right) \hat{w}_2^2-\frac{1}{2}\left(\lambda _5+\lambda _6\right) \hat{w}_S^2,\\
\lambda_4=0, \quad \lambda _7= -\frac{  \sin \left(2 \left(\sigma _1-\sigma _2\right)\right) \hat{w}_2^2}{\sin 2 \sigma _1\hat{w}_S^2}\left(\lambda _2+\lambda _3\right).
\end{gather}

Following the approach of Ref.~\cite{Ogreid:2017alh}, we transform to the Higgs basis:
\begin{equation}
\left( \begin{array}{c}
h_1^\prime \\
h_2^\prime \\
h_S^\prime \\
\end{array}  \right) = \left( \begin{array}{ccc} 
\frac{1}{N_1} (\hat {w_1}  & \hat {w_2} & \hat {w_S})  \\
\frac{1}{N_2} (\hat {w_2} & - \hat {w_1}& 0) \\ 
\frac{1}{N_3} ( \hat {w_1}  &  \hat {w_2} & X) \\
\end{array} \right) \left( \begin{array}{ccc} 
e^{-i\sigma_1} & 0 & 0 \\
0 & e^{-i\sigma_2}  & 0 \\
0 & 0 & 1 
\end{array} \right) 
 \left( \begin{array}{c}
h_1 \\
h_2 \\
h_S \\
\end{array}  \right),
\label{basis3}
\end{equation}
where $X$ is chosen to make lines 1 and 3 orthogonal, and with
$N_1,\, N_2,\, N_3$
normalization factors.
Performing this transformation, and making use of the freedom to perform an overall phase transformation on all fields, we see that the potential remains real, showing that CP is conserved.

In this example, the matrix $U$ of Eq.~(\ref{Eq:U-spont}) is built by following two of the
steps described for the vacuum C-III-c.  The first step, an $SO(2)$ rotation of 
$h_1$ and $h_2$, corresponds to a symmetry of the Lagrangian because this vacuum also
requires $\lambda_4 = 0$.
In this case, the moduli of $w_1$ and $w_2$ are related, see Eq.~(\ref{Eq:C-IV-e-vacuum}). 
As a consequence, rotating the fields and vacuum into
\begin{equation}
\left(be^{i\gamma_1},be^{i\gamma_2},\hat w_S\right),
\end{equation}
leads to
\begin{equation}
\gamma_1+\gamma_2=0,
\end{equation}
and an overall phase rotation is not needed in order to obtain symmetric phases for
$h_1$ and $h_2$. Although the potential is symmetric under an overall phase rotation,
such a rotation would make the vev of $h_S$ complex and would prevent 
Eq.~(\ref{Eq:U-spont}) from being verified. The second step relies now on the
symmetry for the interchange of $h_1$ and $h_2$. 

The new matrix $U$ thus built allows for  Eq.~(\ref{Eq:U-spont}) to be verified and once 
again we can conclude that CP is not spontaneously broken.  

\section{Summary}
We have discussed powerful methods that exist to check for CP conservation in multi-Higgs-doublet models, NHDM:
\begin{itemize}
\item
For $N=2$, a full set of CP-odd invariants are established that can reveal whether or not CP is conserved.
\item
For $N\geq3$ the analysis becomes more complicated and we advocate going first to the Higgs basis and checking
whether or not the remaining $U(N-1)$ rotation freedom allows to transform into a potential with real 
coefficients.
\end{itemize}
\bigskip

\noindent
{\bf Acknowledgements: }
We thank the local organizers of Corfu 2017 for the very fruitful scientific meeting and the warm hospitality.
The work of PO was supported in part by the Research Council of Norway.
The work of MNR was partially supported by Funda\c{c}\~ao
para a Ci\^encia e a Tecnologia (FCT, Portugal)
through the projects CERN/FIS-NUC/0010/2015, and
CFTP-FCT Unit 777 (UID/FIS/00777/2013) which are partially
funded through POCTI (FEDER), COMPETE, QREN and EU. 
MNR benefited from COST support for a STSM to visit the University of Bergen under COST action CA15108
and also benefited from discussions that took place at the University of Warsaw 
during visits supported by the the HARMONIA project of the National Science Centre, Poland, under 
contract UMO-2015/18/M/ST2/00518 (2016-2019).


\begin{thebibliography}{99}

\bibitem{Gunion:1989we}
  J.~F.~Gunion, H.~E.~Haber, G.~L.~Kane and S.~Dawson,
  ``The Higgs Hunter's Guide,''
  Front.\ Phys.\  {\bf 80} (2000) 1.
  
\bibitem{Branco:2011iw}
  G.~C.~Branco, P.~M.~Ferreira, L.~Lavoura, M.~N.~Rebelo, M.~Sher and J.~P.~Silva,
  ``Theory and phenomenology of two-Higgs-doublet models,''
  Phys.\ Rept.\  {\bf 516} (2012) 1
  doi:10.1016/j.physrep.2012.02.002
  [arXiv:1106.0034 [hep-ph]].
  
\bibitem{Ivanov:2017dad}
  I.~P.~Ivanov,
  ``Building and testing models with extended Higgs sectors,''
  Prog.\ Part.\ Nucl.\ Phys.\  {\bf 95} (2017) 160
  doi:10.1016/j.ppnp.2017.03.001
  [arXiv:1702.03776 [hep-ph]].
  
\bibitem{Olaussen:2010aq}
  K.~Olaussen, P.~Osland and M.~A.~Solberg,
  ``Symmetry and Mass Degeneration in Multi-Higgs-Doublet Models,''
  JHEP {\bf 1107} (2011) 020
  doi:10.1007/JHEP07(2011)020
  [arXiv:1007.1424 [hep-ph]].

\bibitem{Grimus:1995zi}
  W.~Grimus and M.~N.~Rebelo,
  ``Automorphisms in gauge theories and the definition of CP and P,''
  Phys.\ Rept.\  {\bf 281} (1997) 239
  doi:10.1016/S0370-1573(96)00030-0
  [hep-ph/9506272].

\bibitem{Branco:1983tn}
  G.~C.~Branco, J.~M.~Gerard and W.~Grimus,
  ``Geometrical T Violation,''
  Phys.\ Lett.\  {\bf 136B} (1984) 383.
  doi:10.1016/0370-2693(84)92024-0

  
\bibitem{Branco:1999fs} 
G.~C.~Branco, L.~Lavoura and J.~P.~Silva, 
``CP Violation,'' Int.\ Ser.\ Monogr.\ Phys.\ \textbf{103} Oxford University
Press (1999). 

\bibitem{Branco:2005em}
  G.~C.~Branco, M.~N.~Rebelo and J.~I.~Silva-Marcos,
  ``CP-odd invariants in models with several Higgs doublets,''
  Phys.\ Lett.\ B {\bf 614} (2005) 187
  doi:10.1016/j.physletb.2005.03.075
  [hep-ph/0502118].
 
\bibitem{Gunion:2005ja}
  J.~F.~Gunion and H.~E.~Haber,
  ``Conditions for CP-violation in the general two-Higgs-doublet model,''
  Phys.\ Rev.\ D {\bf 72} (2005) 095002
  doi:10.1103/PhysRevD.72.095002
  [hep-ph/0506227].

\bibitem{Grzadkowski:2016szj}
  B.~Grzadkowski, O.~M.~Ogreid and P.~Osland,
  ``Spontaneous CP violation in the 2HDM: physical conditions and the alignment limit,''
  Phys.\ Rev.\ D {\bf 94} (2016) no.11,  115002
  doi:10.1103/PhysRevD.94.115002
  [arXiv:1609.04764 [hep-ph]].

\bibitem{Lavoura:1994fv}
  L.~Lavoura and J.~P.~Silva,
  ``Fundamental CP violating quantities in a SU(2) x U(1) model with many Higgs doublets,''
  Phys.\ Rev.\ D {\bf 50} (1994) 4619
  doi:10.1103/PhysRevD.50.4619
  [hep-ph/9404276].

\bibitem{Botella:1994cs}
  F.~J.~Botella and J.~P.~Silva,
  ``Jarlskog - like invariants for theories with scalars and fermions,''
  Phys.\ Rev.\ D {\bf 51} (1995) 3870
  doi:10.1103/PhysRevD.51.3870
  [hep-ph/9411288].

\bibitem{Grzadkowski:2014ada}
  B.~Grzadkowski, O.~M.~Ogreid and P.~Osland,
  ``Measuring CP violation in Two-Higgs-Doublet models in light of the LHC Higgs data,''
  JHEP {\bf 1411} (2014) 084
  doi:10.1007/JHEP11(2014)084
  [arXiv:1409.7265 [hep-ph]].

\bibitem{Grzadkowski:2015zma}
  B.~Grzadkowski, O.~M.~Ogreid and P.~Osland,
  ``Testing the presence of CP violation in the 2HDM,''
  PoS CORFU {\bf 2014} (2015) 086
  [arXiv:1504.06076 [hep-ph]].

\bibitem{Varzielas:2016zjc}
  I.~de Medeiros Varzielas, S.~F.~King, C.~Luhn and T.~Neder,
  ``CP-odd invariants for multi-Higgs models: applications with discrete symmetry,''
  Phys.\ Rev.\ D {\bf 94} (2016) no.5,  056007
  doi:10.1103/PhysRevD.94.056007
  [arXiv:1603.06942 [hep-ph]].


\bibitem{deMedeirosVarzielas:2017ote}
  I.~de Medeiros Varzielas, S.~F.~King, C.~Luhn and T.~Neder,
  ``Spontaneous CP violation in multi-Higgs potentials with triplets of $\Delta(3n^2)$ and $\Delta(6n^2)$,''
  JHEP {\bf 1711} (2017) 136
  doi:10.1007/JHEP11(2017)136
  [arXiv:1706.07606 [hep-ph]].

\bibitem{Davidson:2005cw}
  S.~Davidson and H.~E.~Haber,
  ``Basis-independent methods for the two-Higgs-doublet model,''
  Phys.\ Rev.\ D {\bf 72} (2005) 035004
   Erratum: [Phys.\ Rev.\ D {\bf 72} (2005) 099902]
  doi:10.1103/PhysRevD.72.099902, 10.1103/PhysRevD.72.035004
  [hep-ph/0504050].


\bibitem{Ogreid:2017alh}
  O.~M.~Ogreid, P.~Osland and M.~N.~Rebelo,
  ``A Simple Method to detect spontaneous CP Violation in multi-Higgs models,''
  JHEP {\bf 1708} (2017) 005
  doi:10.1007/JHEP08(2017)005
  [arXiv:1701.04768 [hep-ph]].


\bibitem{Donoghue:1978cj}
  J.~F.~Donoghue and L.~F.~Li,
  ``Properties of Charged Higgs Bosons,''
  Phys.\ Rev.\ D {\bf 19} (1979) 945.
  doi:10.1103/PhysRevD.19.945

\bibitem{Georgi:1978ri}
  H.~Georgi and D.~V.~Nanopoulos,
  ``Suppression of Flavor Changing Effects From Neutral Spinless Meson Exchange in Gauge Theories,''
  Phys.\ Lett.\  {\bf 82B} (1979) 95.
  doi:10.1016/0370-2693(79)90433-7
  
\bibitem{Lee:1973iz}
  T.~D.~Lee,
  ``A Theory of Spontaneous T Violation,''
  Phys.\ Rev.\ D {\bf 8} (1973) 1226.
  doi:10.1103/PhysRevD.8.1226

\bibitem{Branco:1985aq}
  G.~C.~Branco and M.~N.~Rebelo,
  ``The Higgs Mass in a Model With Two Scalar Doublets and Spontaneous {CP} Violation,''
  Phys.\ Lett.\  {\bf 160B} (1985) 117.
  doi:10.1016/0370-2693(85)91476-5


\bibitem{Gunion:2002zf}
  J.~F.~Gunion and H.~E.~Haber,
  ``The CP conserving two Higgs doublet model: The Approach to the decoupling limit,''
  Phys.\ Rev.\ D {\bf 67} (2003) 075019
  doi:10.1103/PhysRevD.67.075019
  [hep-ph/0207010].
  

\bibitem{Derman:1978rx}
  E.~Derman,
  ``Flavor Unification, $\tau$ Decay and $b$ Decay Within the Six Quark Six Lepton {Weinberg-Salam} Model,''
  Phys.\ Rev.\ D {\bf 19} (1979) 317.
  doi:10.1103/PhysRevD.19.317
 
\bibitem{Das:2014fea}
  D.~Das and U.~K.~Dey,
  ``Analysis of an extended scalar sector with $S_3$ symmetry,''
  Phys.\ Rev.\ D {\bf 89} (2014) no.9,  095025
   Erratum: [Phys.\ Rev.\ D {\bf 91} (2015) no.3,  039905]
  doi:10.1103/PhysRevD.91.039905, 10.1103/PhysRevD.89.095025
  [arXiv:1404.2491 [hep-ph]].

\bibitem{Emmanuel-Costa:2016vej}
  D.~Emmanuel-Costa, O.~M.~Ogreid, P.~Osland and M.~N.~Rebelo,
  ``Spontaneous symmetry breaking in the $S_3$-symmetric scalar sector,''
  JHEP {\bf 1602} (2016) 154
   Erratum: [JHEP {\bf 1608} (2016) 169]
  doi:10.1007/JHEP08(2016)169, 10.1007/JHEP02(2016)154
  [arXiv:1601.04654 [hep-ph]]; \\
%
  D.~Emmanuel-Costa, O.~M.~Ogreid, P.~Osland and M.~N.~Rebelo,
  ``CP Violation in the scalar sector,''
  PoS CORFU {\bf 2015} (2016) 044
  [arXiv:1604.00637 [hep-ph]].

\end{thebibliography}
\end{document}